\documentclass[a4paper,11pt]{article}
\usepackage{jinstpub} 
\usepackage{lineno}

\usepackage[utf8]{inputenc}
\usepackage[T1]{fontenc}
\usepackage{subcaption}
\def \Li {${}^8{\rm Li}$ }

\begin{document}
\title{Neutrino yield and neutron shielding calculations for
a high-power target installed in an underground setting}

\author[a]{Adriana~Bungau} 
\author[a]{Jose~Alonso}
\author[b,1]{Roger~Barlow\note{Corresponding Author}}
\author[c]{Larry~Bartoszek}
\author[a]{Janet~Conrad}\author[d]{Michael~Shaevitz}\author[e]{        Joshua~Spitz}
\author[a]{Daniel~Winklehner}
 
\affiliation[a]{
Massachusetts Institute of Technology, 77 Massachusetts Ave, Cambridge, MA 02139, USA}%
\affiliation[b]{The University of Huddersfield, Huddersfield HD1 3DH, UK}
\affiliation[c]{Bartoszek Engineering, Aurora IL, USA}
\affiliation[d]{ 
Columbia University, New York, New York 10027, USA}
\affiliation[e]{University of Michigan, Ann Arbor, Michigan 48109, USA}

\emailAdd{Roger.Barlow@hud.ac.uk}



\arxivnumber{2409.10211} 

\abstract{
With the ever increasing beam power at particle accelerator-based facilities 
for nuclear and particle physics, radioactive isotope production, and 
nuclear engineering, it becomes increasingly important to have targets that can withstand this power, and shielding
to block the secondary particles produced.
Here we present Monte Carlo (MC) calculations using the well-established Geant4 software
to 
predict the antineutrino yield of a $^8$Li Decay-At-Rest (DAR) source. The source relies on 600~kW of beam power from a continuous wave proton beam impinging on a beryllium 
target, where spallation neutrons are captured by $^7$Li to produce the $^8$Li. 
We further present an in-depth treatment of the neutron shielding surrounding 
this target.
We show that we can produce the high antineutrino flux needed for the discovery-level
experiment IsoDAR, searching for ``sterile'' neutrinos (predicted new fundamental particles) and other beyond standard model physics,
while maintaining a neutron flux in the detector that is below natural backgrounds.
The methods presented in this paper are easily transferable to other high-power
targets and their associated shielding.
}

\maketitle

\flushbottom

\section{Introduction \label{sec:intro}}

Underground laboratories provide areas of exceedingly low background for very sensitive experiments~\cite{Formaggio:Background}. 
 Placing these experiments in caverns with typically a thousand meters of rock overburden reduces the cosmic muon flux by a factor of several million compared to the rate at the surface.
 For experiments that cannot make use of a very narrow beam time window to minimize the effects of atmospheric muons, this shielding is absolutely necessary.
An example of such an experiment is IsoDAR~\cite{alonsoNeutrinoPhysicsOpportunities2022,winklehnerIsoDARYemilabDefinitive2023}, 
designed to study neutrinos and search for rare phenomena.  
The configuration of this experiment places an intense neutrino source in close proximity to a large liquid scintillation detector.  
Extremely high fluxes of neutrons and other radiation are produced in the vicinity of the target, 
 a necessary consequence of the reactions needed to produce the desired neutrino flux.

This poses the question whether it is possible to deploy such a neutrino production facility in an underground laboratory while maintaining the quiet environment for other experiments in the same facility.
The answer is yes, but it must be done with careful attention to shielding, regulatory requirements, and containment of all sources of radiation and neutrons. As explained later, this implies fluxes that produce rock activation below 10 Bq/g (for Korea, or 0.1 Bq/g for Japan),  and very small numbers of neutron-induced background events in the detector.

Strategies for achieving this have been presented in several previous papers ~\cite{alonsoIsoDARYemilabReport2022,Shieldingpaper,Bungau_2019}, which addressed the deployment of IsoDAR at the Kamioka Observatory, in close proximity to the KamLAND liquid scintillation detector.  The present work focuses on a new site, Yemilab, in South Korea, and presents designs appropriate for this site and the planned $\nu$EYE 2.4 kiloton liquid scintillator detector (formerly known as the Liquid Scintillator Counter, or LSC). More importantly, this paper addresses considerations and general techniques that should be employed in matching such a high-radiation source with the clean and low-background underground environment.

The IsoDAR experiment will use a 60~MeV, 10 mA proton beam,  produced by a novel compact cyclotron~\cite{winklehnerNewFamilyHighcurrent2024, winklehnerOrderofmagnitudeBeamCurrent2022}, directed onto a $^9{\rm Be}$ target to produce neutrons, through reactions such as beryllium break-up. These neutrons
enter a surrounding sleeve of beryllium, for additional neutron production, and $^7{\rm Li}$. Neutron capture on  $^7{\rm Li}$ results in $^8{\rm Li}$, which rapidly $\beta$ decays ($t_{1/2}=0.839$~s) to provide a source of electron antineutrinos ($\overline{\nu}_e$):
\begin{align*}
    p + ^9{\rm Be} &\to \alpha+\alpha+n + p\\
    n+^7{\rm Li} &\to ^8{\rm Li}\to ^8{\rm Be}+e^-+\overline{\nu}_e
\end{align*}
The $\overline{\nu}_e$ can be detected through inverse beta decay (IBD), $\overline{\nu}_e + p \to e^+ + n$, in a large liquid scintillator detector (e.g. the $\nu$EYE at Yemilab~\cite{Seo:2023xku})
enabling a sensitive search for neutrino oscillations at short distances, which will 
definitively elucidate the hints (from, e.g., radioactive-source-based~\cite{gallex, sage, Barinov:2022wfh} and accelerator-based short baseline neutrino experiments~\cite{lsnd3,MiniBooNENeutrino2021}) possibly indicating the existence of ``sterile'' neutrinos,
and allow a range of other beyond-standard-model (BSM) physics searches~\cite{alonsoNeutrinoPhysicsOpportunities2022,Waites:2022tov,Hostert:2022ntu}.
\begin{figure}[b!]
    \centering
    \includegraphics[width=1.0\columnwidth]{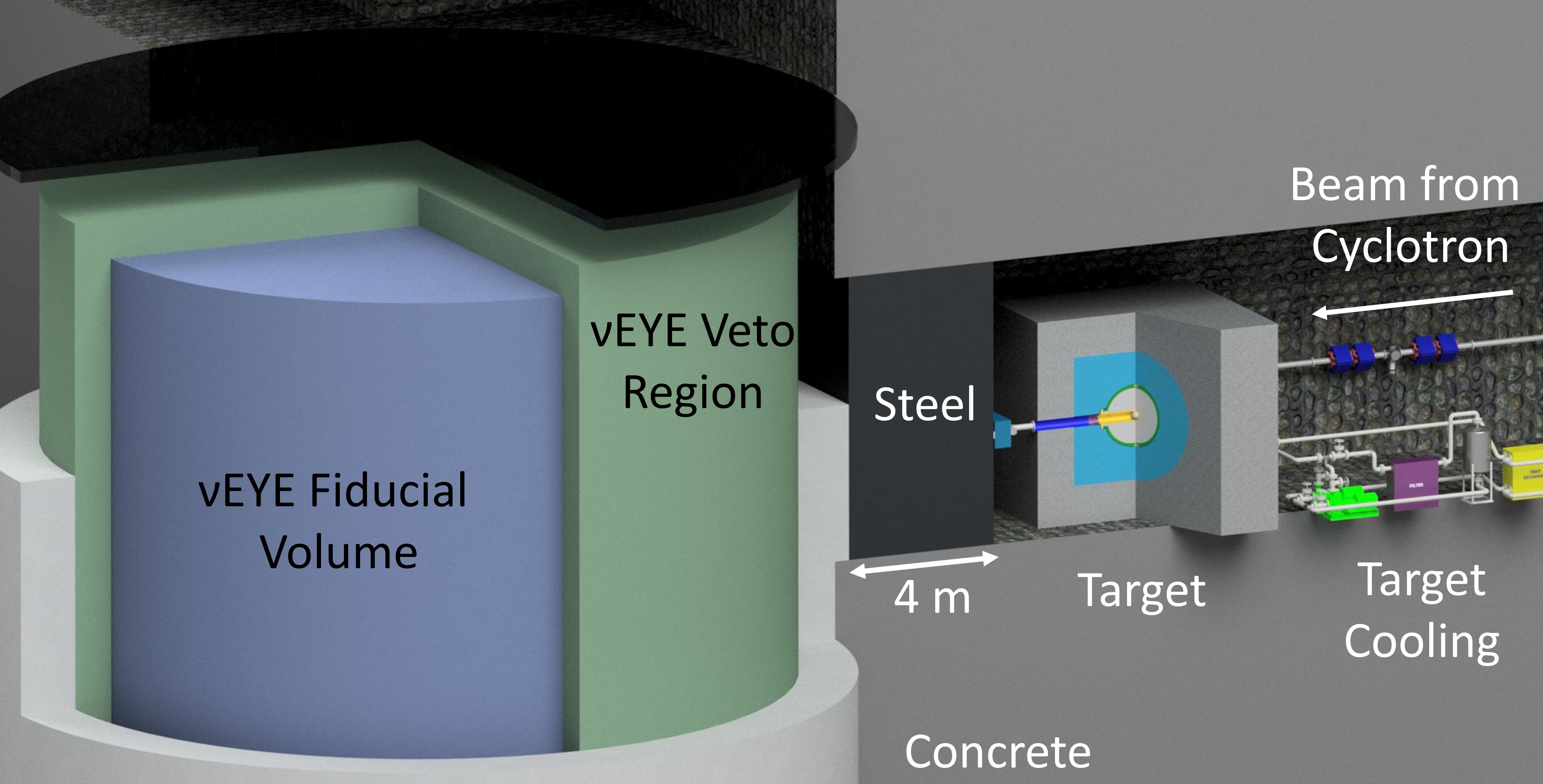}
    \caption{The layout of the IsoDAR target and the $\nu$EYE detector at Yemilab~\cite{Seo:2023xku}.\label{fig:layout}}
\end{figure}
The proposed design calls for five years of running, giving a data sample of approximately 1.67 million selected IBD events.
This requires the production of  $4.6 \times 10^{15}$ antineutrinos/second. Optimising this number required careful
consideration of the target composition and geometry, described in  \cite{Bungau_2019} and references therein, and the study of the neutron and antineutrino yield is the first topic of this paper.
The second topic is neutron shielding; the flux within the target is, by design, extremely high and shielding is required to screen the surroundings, both for reasons of safety and to manage
backgrounds in the $\nu$EYE.

An $E_{vis}>3$ MeV requirement on candidate single-flash-signal events in the fiducial volume, which starts around 3 m from the outer wall of the detector, is planned. 
Around $E_{vis} = 3$ MeV, the background behavior changes\cite{Formaggio:Background,Abe,Suekane}. Below 3 MeV,
there are large number of backgrounds from radioactive decays in the
PMTs and the inner detector tank, such as ${}^{208} {\rm Tl}$ and ${}^{40}{\rm K}$. On
the other hand, contributions from radioactivity are small above
3 MeV (see Ref.~\cite{alonsoNeutrinoPhysicsOpportunities2022} for a detailed analysis).
This requirement was motivated by the goal of reducing backgrounds from natural radioactivity to an acceptable level, and still allows 0.07 Hz of inverse beta decay interactions and 0.2 mHz of elastic scattering events to be accepted for physics studies.    Although introduced for other reasons, this cut has the benefit of also eliminating 
neutrons from the target that 
penetrate the outer regions of the detector, but then deposit less than 3 MeV of energy---hence this discussion focuses on 
neutrons above 3 MeV, aiming for a rate of $< 1 \mu$Hz given the shielding we describe, which is three orders of magnitude below the rate for elastic scatters, making this background negligible for neutrino analyses and allowing high sensitivity for Beyond Standard Model event searches.


Both optimising antineutrino production and shielding have been considered in two previous publications\cite{Bungau_2019,Shieldingpaper}.
An update is necessary for several reasons.
\begin{enumerate}
    \item The target has been redesigned to have three hemispherical beryllium shells rather than a simple disc, for relieving unacceptably high thermal stresses at the edges of the original flat target.
    \item The shape and dimensions of the surrounding lithium-beryllium sleeve have been optimised to maximise antineutrino production with the most efficient utilization of the highly-enriched $^7$Li inventory. 
    \item The previous accounts considered the KamLAND
site, and here we address requirements for Yemilab. ( However the methods employed can be applied to any site,
depending on physical and regulatory requirements.)
    \item Various other relevant studies have been done in response to questions that have arisen.
\end{enumerate}

\begin{figure}[t!]
    \centering
    \includegraphics[height=4.25cm]{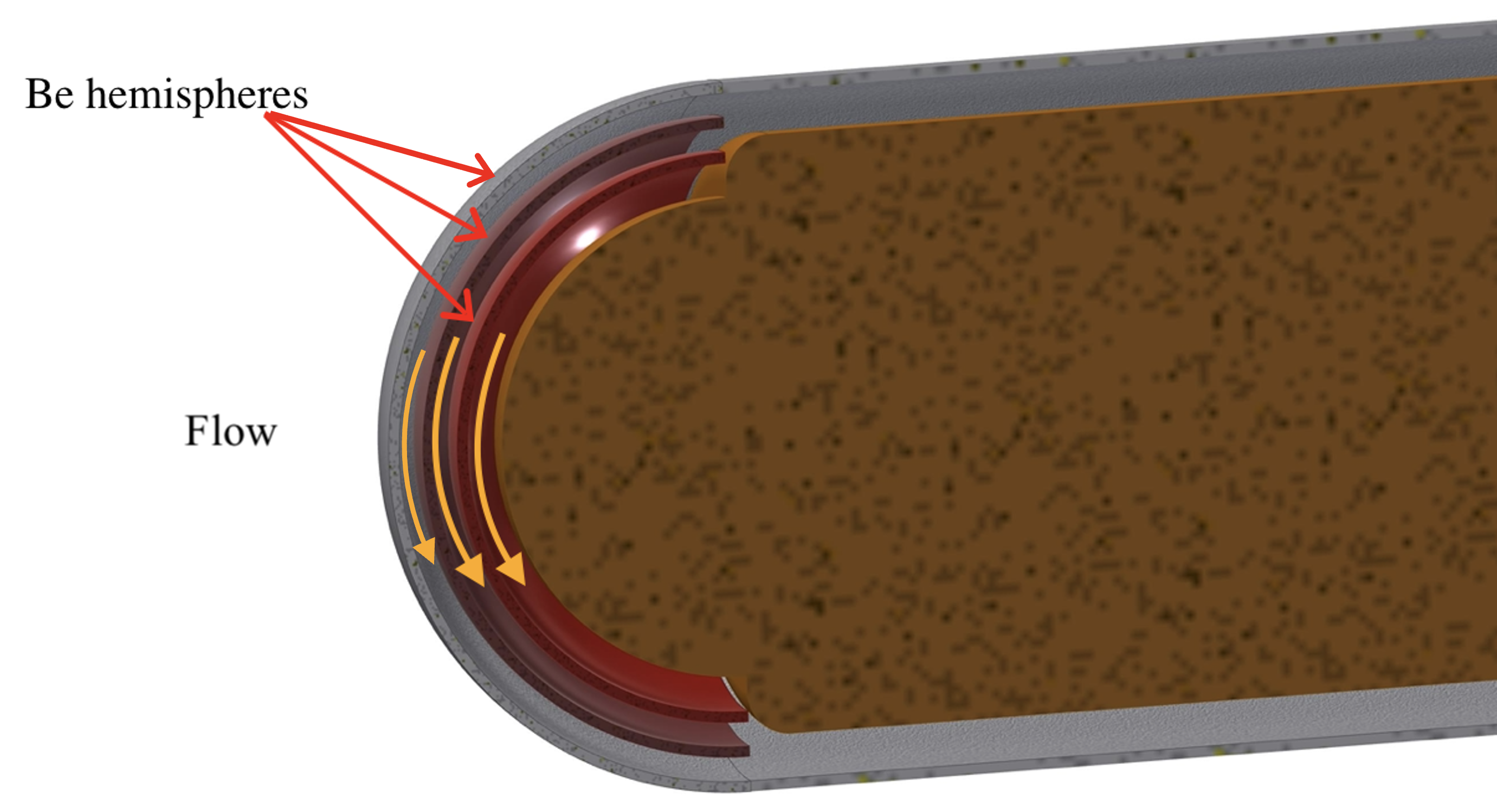}
    \caption{The target: beryllium shells and cooling heavy water.}
    \label{fig:target}
\end{figure}

The general layout of the experiment is shown in Figure~\ref{fig:layout}. The beam from the cyclotron (off the figure and about 20 meters to the right, in an existing cavern specifically constructed for it by the Yemilab Laboratory) is bent through $180^\circ$ before striking the target. 
The target assembly, beryllium hemispheres and cooling circuits shown in Figure~\ref{fig:target} -- referred to as the "torpedo" -- is inserted through the downstream end of the beam pipe.  The outer beryllium hemisphere is 20 cm diameter, the beam is 
spread out to distribute
the beam power over the entire target surface.  This maintains the maximum power density at less than 
8 kW/cm$^2$, a safe limit for cooling design.  The cart with pumps and heat exchangers is seen behind the shielding in Figure~\ref{fig:layout}.
This geometry has the advantage of providing space to enable the target assembly to be withdrawn and replaced from the downstream end of the beam pipe with remote handlers, which we anticipate will be necessary several times during the lifetime of the experiment. Boreholes drilled into the cavern walls serve as storage for new and spent targets. As the proton beam is thus directed away from the $\nu$EYE it has the additional benefit of reducing the background from high energy neutrons directed towards the detector, as will be discussed later  (see Fig.~\ref{fig:n-out-target}). The $\nu$EYE is shielded from the target by steel and concrete: the thickness of these layers is an important design choice, as increasing the thickness reduces the neutron background but also the antineutrino signal 
(since the antineutrino yield is isotropic, increasing shielding increases the distance from the detector, and  the flux drops as 1/r$^2$).
In the present design 
the center of the fiducial volume of the $\nu$EYE is 17.6 meters from the target, including a 4-meter slab of steel and the beamline magnet bringing the beam into the target.

The IsoDAR target design presented in Ref.~\cite{Bungau_2019} and updated in Ref.~\cite{alonsoIsoDARYemilabReport2022} has been redesigned, as shown in Figure~\ref{fig:target}.

At the vacuum interface, 
the proton beam impinges on a 3~mm thick hemispherical
beryllium shell. Behind it lie two more concentric, 3~mm thick, beryllium shells
separated by 7~mm deep channels of heavy water, pumped and cooled by a remote external circulator. Finally, the centre is comprised of
a solid beryllium block.
The design contains 480 kg of ${}^7{\rm Li}$ and 1437 kg of Be in the sleeve, another  44.5 kg of Be in the target, 
and 18.65 kg of heavy water.
Full details, including strategies to avoid stopping of protons in beryllium layers of the target, which is known to lead to embrittlement because of poor solubility of hydrogen in beryllium~\cite{Rinckel_2012}, will be given in a projected future publication
on the mechanical design and fluid dynamics of the target and cooling.

\begin{figure}[t!]
    \centering
    \includegraphics[height=4.25cm]{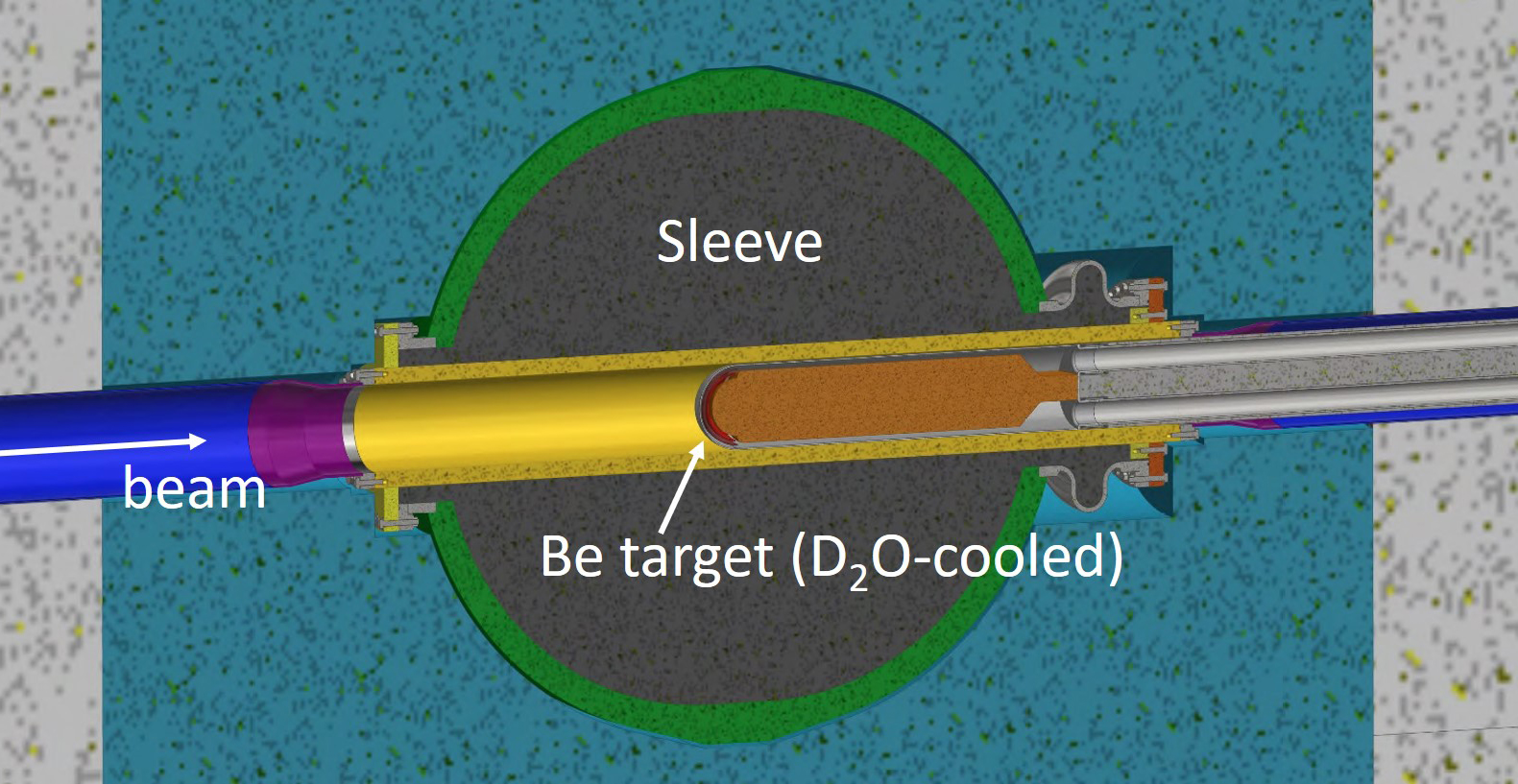}
    \caption{The target surrounded by the Li-Be sleeve.}
    \label{fig:sleeve}
\end{figure}

The cylindrical Li/Be sleeve of the previous design has been replaced by a (roughly) spherical section, as shown in Figure~\ref{fig:sleeve}, following the optimisation analysis developed in~\cite{Bungau_2019}. 

\section{Methods}

To simulate all stages of the process we use the 
   Geant4 
code\cite{agostinelli,ALLISON2006,ALLISON2016}. 
Originally developed for the description of particle physics detectors, it has since been
expanded and applied to many fields. We use version Geant4-11.0.2. 

Different stages of the simulation require different Geant4 physics lists, as no single option is suitable for all processes involved. For the initial stage of neutron and \Li production, we employed QGSP\_BIC\_AllHP, which combines the binary cascade model for higher energies with high-precision, data-driven models (ParticleHP) for low-energy proton and neutron interactions based on ENDF and TALYS cross-section libraries. This physics list is recommended for applications such as BNCT (Boron Neutron Capture Therapy) and provides reliable agreement with experimental data.

For the shielding studies, we used the Shielding physics list \cite{Shieldinglist}. Originally developed for neutron penetration and ion–ion collision studies, it is also applicable to calorimetry and low-background experiments. Its high-energy treatment derives from FTFP\_BERT, with radioactive decay processes included to account for background radiation, while retaining the same evaluated neutron cross-section libraries as the high-precision models.
For energies above 20 MeV, the Bertini cascade inelastic scattering model~\cite{Wright} is used for energies up to 3 GeV with parameters tuned with the help of the JENDL HE-2007 neutron cross sections~\cite{JENDL}  available at these higher energies. 

In order to optimise the production of antineutrinos we must maximise that of the precursor
neutrons. The disintegration of $^9$Be was the method by which the neutron was first seen\cite{Chadwick} and is still the
most efficient way of producing neutrons using a small accelerator. (Lithium also gives neutrons, but is structurally unworkable, with  a low melting point.)


The target constitutes a very high intensity neutron source, and shielding
must be put in place to attenuate these penetrating particles beyond the production sleeve. The IsoDAR experiment will be deep underground and will be controlled remotely, so exposure of personnel is not an issue. Components will become activated, but maintenance
work will be possible after a suitable cooling-off period. The two principal concerns are for the activation in the surrounding rock, which must remain below a statutorily-required level (discussed later) on completion of the experiment, and backgrounds in the $\nu$EYE detector, where a high energy neutron interaction, with possible subsequent capture, can 
mimic the double-coincidence signature of an antineutrino
event.  

The shielding materials considered were:
\begin{itemize}
\item
Iron, which has a large interaction cross section for high energy neutrons (above $\sim 2$ MeV).

\item
Boron loaded concrete, which has a high absorption cross section for slow neutrons.
\end{itemize}
A typical configuration considered was a thick layer of iron, which 
softened the spectrum, followed by a layer of boron loaded concrete to remove the resulting slow neutrons. 

For environmental shielding inside the cavern and mitigation of rock activation, attenuation of all neutrons, respective of energy, is important.  For this, it is optimal to provide a combination of iron to efficiently reduce the flux of higher-energy neutrons through inelastic collisions, and absorption of slower neutrons through boron 
loaded concrete.  For backgrounds in $\nu EYE$ we only need consider neutrons above 3 MeV, as discussed in Section~\ref{detector}, so the concrete attenuation is not really required.  But in addition, the buffer and veto layers in the detector provide ample attenuation of neutrons emerging from the large steel block between the target and the detector.


\subsection{Validation}

Although Geant4 is very widely used for simulation in particle physics experiments, its predictions need to be validated and uncertainties estimated, and the impact of such uncertainties on the proposed experiment evaluated.

\subsubsection{Validation of neutron production rates}

\begin{figure}[htbp]
    \centering
    \begin{subfigure}[t]{0.48\textwidth}
        \centering
        \includegraphics[width=\textwidth]{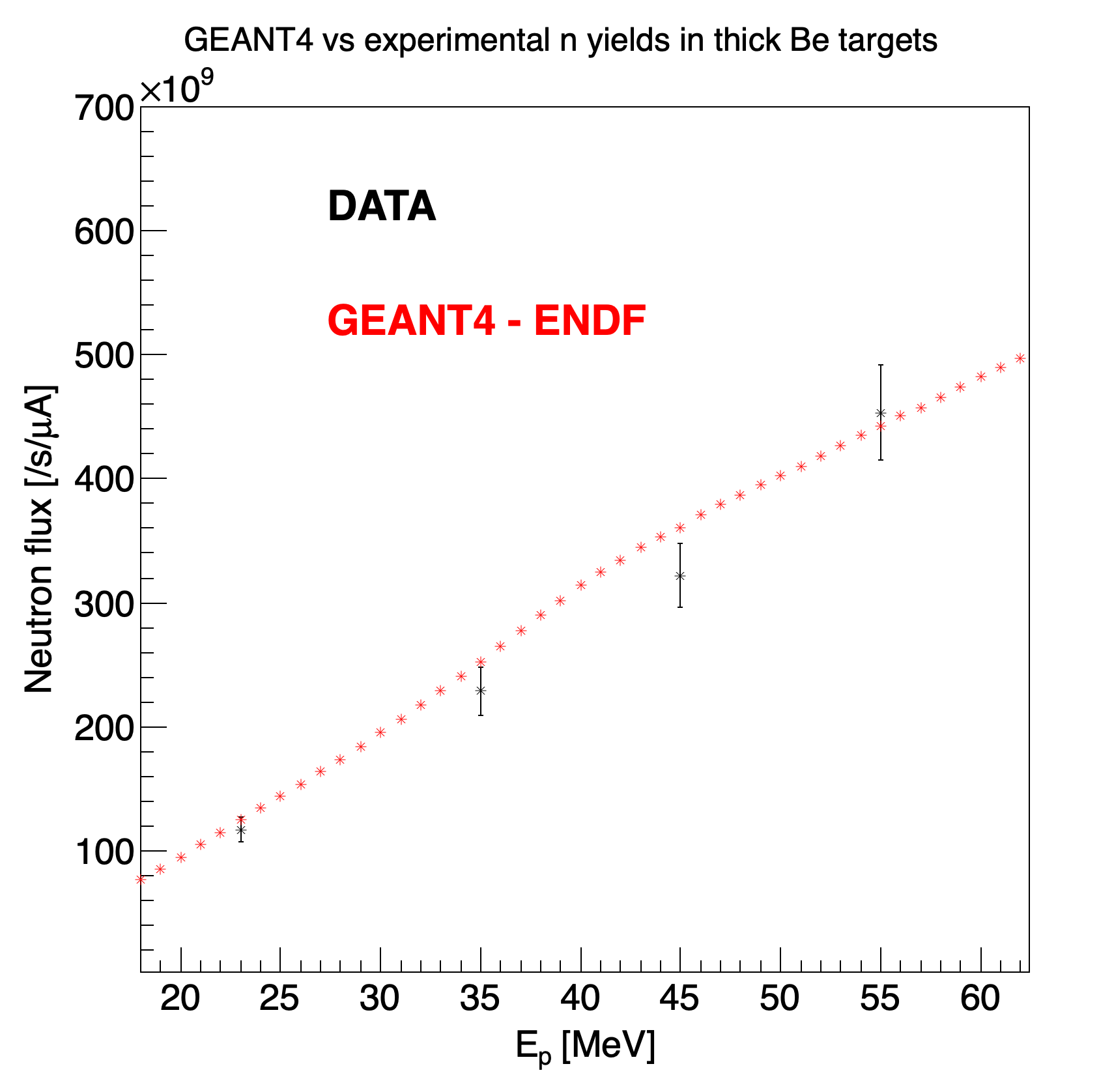}
        \caption{Total neutron yields as a function of proton beam energy: Geant4 reproduces both the magnitude and the beam-energy dependence of the data within ~10\% across the 20–60 MeV range.}
        \label{fig:plot1}
    \end{subfigure}
    \hfill
    \begin{subfigure}[t]{0.48\textwidth}
        \centering
        \includegraphics[width=\textwidth]{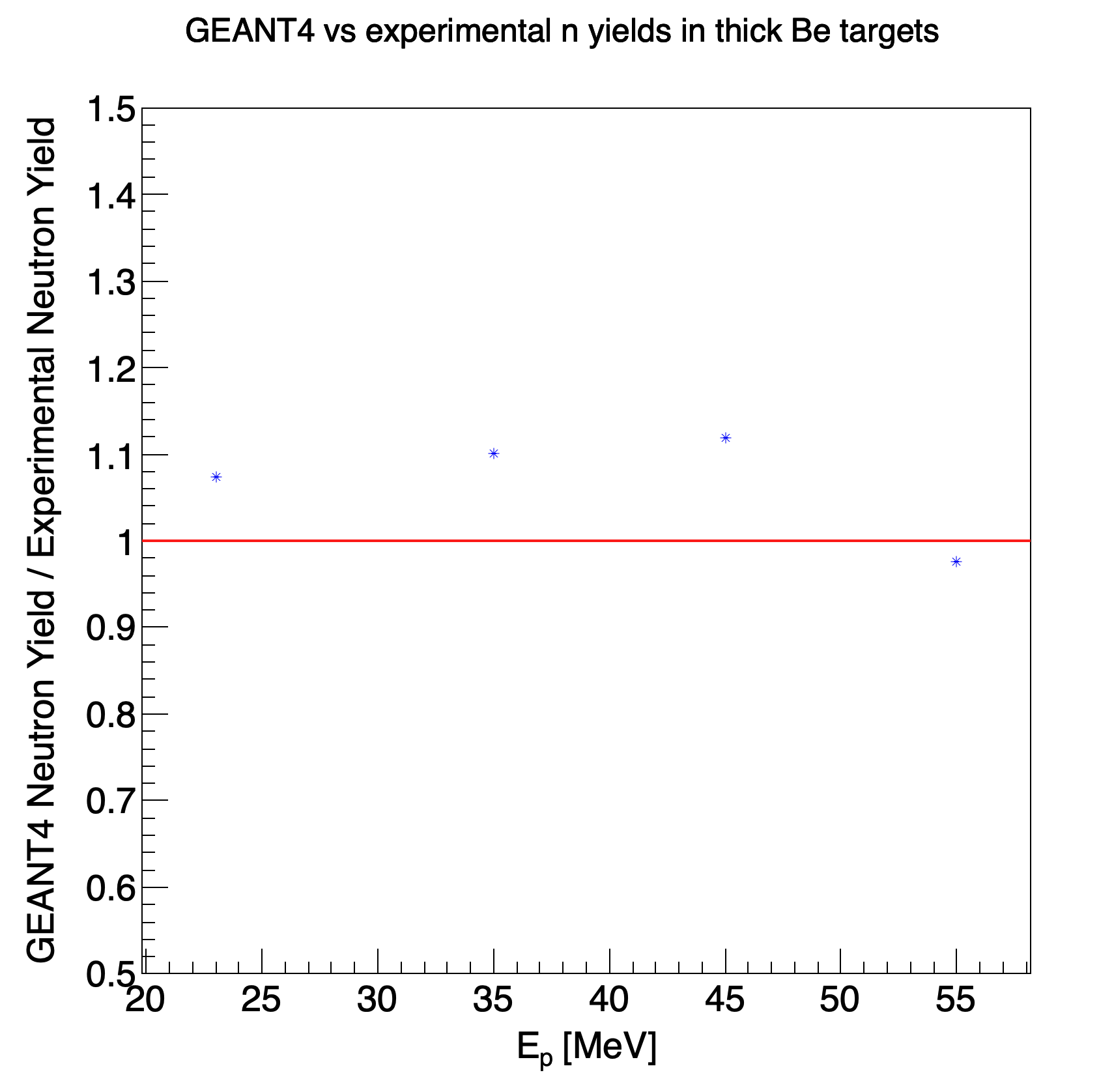}
        \caption{Ratio of Geant4 predictions to experimental measurements, showing values close to unity with a slight overestimation at lower energies and modest underestimation at higher energies.}
        \label{fig:plot2}
    \end{subfigure}
    \caption{Validation of Geant4 predictions against the experimental neutron yield data of Tilquin et al. (2005) for thick Be targets.}
    \label{fig:overall}
\end{figure}

We have conducted a comprehensive validation of the Geant4 physics models relevant to low-energy proton interactions with beryllium 
using experimental data from Osipenko et al.\cite{Osipenko} at 62 MeV and from  Tilquin et al. \cite{tilquin} at 23-80 MeV,
which show good agreement between Geant4 predictions and measured neutron yields and have presented these in previous publications \cite{Shieldingpaper,Bungau_2019}.
We demonstrated that, with appropriate physics models selection, Geant4 accurately reproduces neutron yields for proton energies between 23 MeV and 55 MeV, with only a slight underestimation at higher energies.
 In Fig.~\ref{fig:overall} (not previously pubished) we compare  predicted neutron yields from thick Be targets with  experimental data from \cite{tilquin}.

The good agreement observed provides strong support for using Geant4 to model neutron production in the energy range relevant to our experiment.

If the Geant4 predicted yields do  turn our to be 
optimistic, by 10\% or even higher (though we do not expect that), the impact on the experiment is not catastrophic. 
Lower rates will mean that we need to run the experiment for a longer time or, alternatively, to publish results with somewhat larger statistical uncertainties. 

\subsubsection{Validation of shielding studies}
The Geant4 `Shielding' physics list~\cite{Shieldinglist} has been continuously used in all the shielding aspects of accelerator, target and irradiation facilities workshops~\cite{Hirayama8,Matsuda,Conf1,NEA1,NEA2,NEA3} since 2010 as part of the `Intercomparison of Medium-energy Neutron Attenuation in Iron and Concrete' project ~\cite{Hirayama6}. These studies involved an intercomparison of the available Monte Carlo codes which can be used in shielding calculations, and they show how the Geant4 Shielding physics list has evolved and  improved over the years since it was  added to the comparison list in 2010~\cite{Shieldinglist}. It has been used not only for a direct comparison against other Monte Carlo codes predictions, but also for validation studies in direct comparisons with experimental data, showing very good agreements~\cite{NEA2}. For lower neutron energies, the Shielding physics list uses the high-precision (HP) neutron physics model which includes the available evaluated neutron data for neutron interactions below 20 MeV.



These studies are cited in the present work to clarify that the validation of the relevant Geant4 models has already been completed and demonstrated to be reliable within the energy range of interest.

\section{Results}

\subsection{Neutron and antineutrino yield}

The new IsoDAR design features a hemispherical beryllium target, as described in Section~\ref{sec:intro}.

Neutrons are produced both in the beryllium shells and in the heavy water between them. A pure heavy water target was considered, but
even after optimisation this could produce only half as many neutrons as the beryllium. 
Similarly, cooling with light water produces a significantly lower yield of neutrons.

The target is surrounded by a sleeve, as shown in Figure~\ref{fig:sleeve}. This contains lithium ($^7{\rm Li}$), from which the desired ${}^8$Li is formed, and also beryllium, as this
multiplies the neutron flux 
through the $(n,2n)$ multiplicative reaction.

\begin{figure}[h]
    \centering
    \includegraphics[height=12cm]{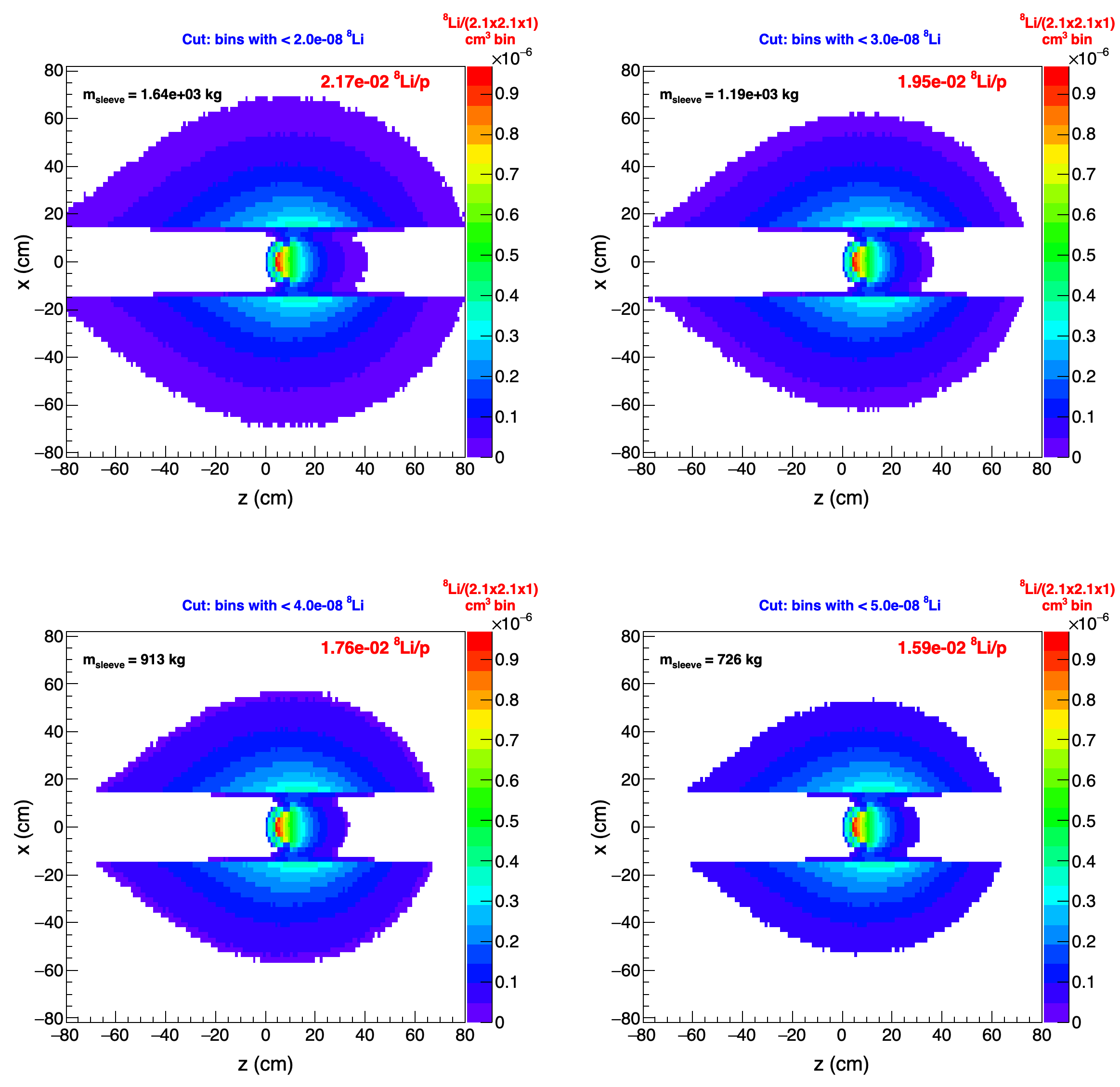}
    \caption{\Li  contour plots inside the target sleeve volume. The beam direction is left to right.}
    \label{fig:contour_plots}
\end{figure}

The previous design~\cite{Bungau_2019} used a cylindrical sleeve, but simulations showed that the neutron
production was, as one would expect, roughly spherical, and so the amount of expensive sleeve material can be reduced by removing
the least productive volumes. An optimization study was performed to determine the best shape and size of the target sleeve to maximise \Li production. The study used a three-dimensional mesh to quantify the spatial distribution of \Li  within the volume of the sleeves. By calculating the \Li  content in each mesh cell, regions that contribute minimally to the overall yield were identified. Mesh cells with \Li concentrations below a predefined threshold were systematically excluded, allowing effective sleeve geometry to be inferred. This approach enables the design of a sleeve that maximizes \Li  production while minimizing unnecessary material, providing a clear guideline for the target geometry in experimental setups. The results are shown in Figure~\ref{fig:contour_plots}.

The optimised design can be approximated by a spherical shape of ~60 cm radius, excluding the outer stainless steel shell.

The previous design studies \cite{Bungau_2019}
showed that the optimum sleeve material was a mixture of beryllium and isotopically pure $^7{\rm Li}$, 
with a Be mass fraction of 75\%. It is formed of 1~mm diameter beryllium spheres 
in lithium.
A FLiBe eutectic mix was also considered, but is less performant~\cite{Bungau_2019}.

\begin{figure}[t!]
    \centering
    \includegraphics[width=1.0\columnwidth]{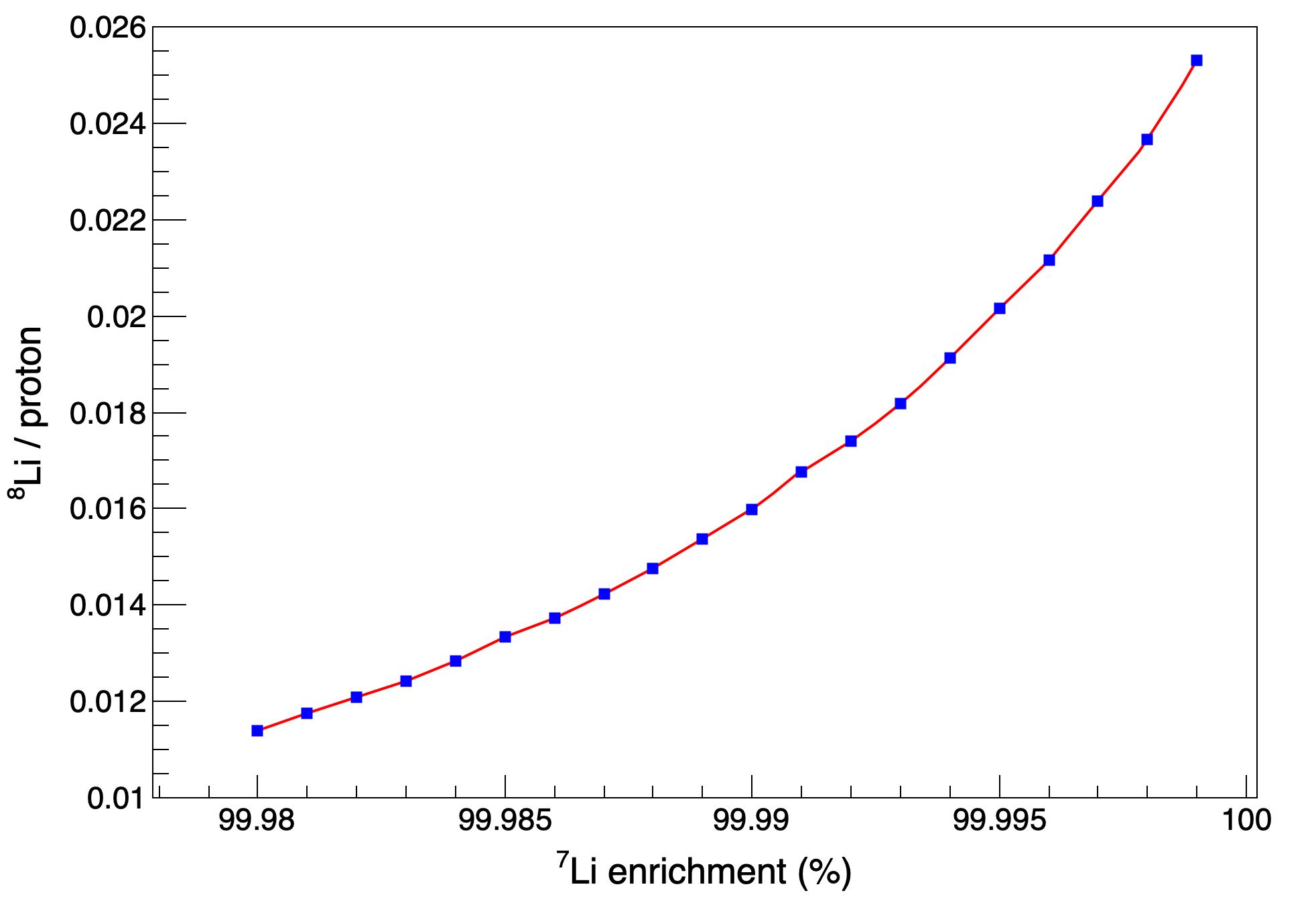}
    \caption{${}^8$Li production as a function of ${}^7$Li enrichment, from a Geant4 simulation.}
    \label{fig:enrichment}
\end{figure}

The previous design achieved a production rate of 0.018 $^8{\rm Li}$ per incident proton.
The new nested-shell target design, necessary for cooling, initially reduced this by a factor $\sim 4$. This figure, which is a key measure of performance, has been increased again by optimisation of the target and sleeve geometry.
We note that ${}^6$Li has a very large neutron cross section,
and even small amounts will have a serious negative effect. Natural lithium is 95.15\% 
pure ${}^7$Li, and we need this to be increased to well over 99\%, as shown in Figure~\ref{fig:enrichment}. A purity of 99.99\% will give 0.016 $^8{\rm Li}$ per proton, which is sufficient for the experiment, and   99.995\% will give 0.020.
Isotopically purified lithium is commercially available due to its importance in molten salt reactors and in the fusion program.
A higher purity would perform even better, but we take the lower number as our baseline,  due to cost and availability. 

${}^8{\rm Li}$ is an ideal single-isotoped neutrino source as the decay produces antineutrinos at very high energy compared to other beta decays.  The IBD process in the liquid scintillator essentially eliminates backgrounds from
neutrons, neutrinos and cosmic rays by requiring two signals in coincidence, from the first flash due to the positron and subsequent second flash from the neutron capture. The IBD reaction requires a neutrino energy of 1.8~MeV, quite high for a beta decay. 
The ${}^8{\rm Li}$ spectrum is peaked at 6 MeV whereas the other isotopes produced (in smaller quantities) in the target produce much lower energies. 
Anti-neutrinos from tritium  decay only go up to 0.019 MeV, from ${}^{10}{\rm Be}$ up to 0.55 MeV.  
There will be a small contribution from ${}^{6}{\rm He}$, which goes up to 3.5 MeV and is produced in very small amounts but otherwise
the antineutrino flux above 2 MeV will only come from ${}^8{\rm Li}$ decay.

Also the ${}^8{\rm Li}$ spectrum is well understood:
it has been extensively studied due to its role in solar neutrino physics, so the systematic uncertainty from the theoretical prediction of the spectrum is very low. It is widely used for tests for BSM physics because the shape of the spectrum is so well known.   See for example\cite{spectrum1,spectrum2}  which describe a test for tensor currents and discusses the SM prediction.

\subsection{Shielding}

\label{sec:backgrounds}

\begin{figure}[b!]
    \centering
    \includegraphics[width=1.0 \columnwidth]{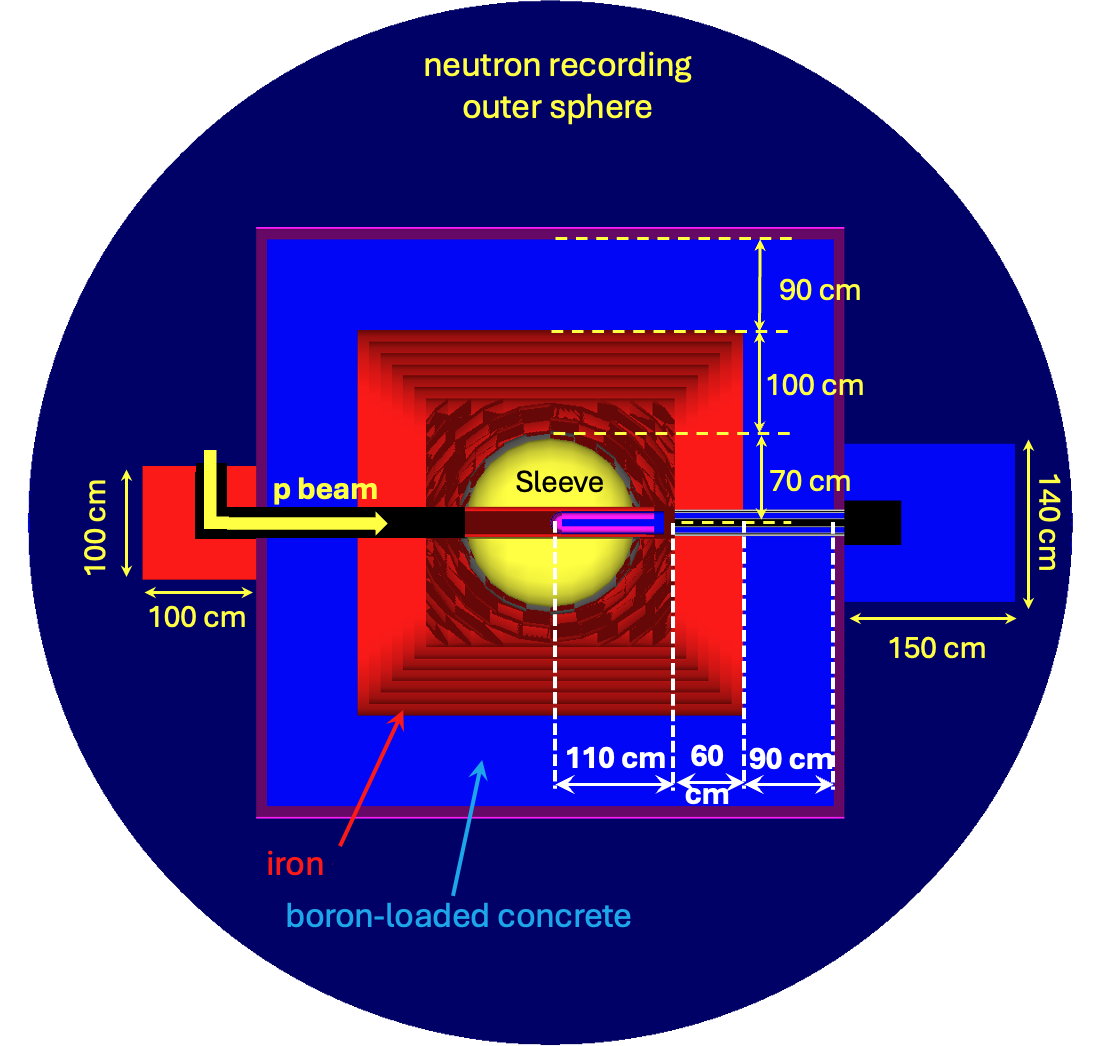}
    \caption{Shielding around the IsoDAR target, shown at the center of the sleeve.}
    \label{fig:shielding}
\end{figure}

Our shielding assessments are made with several goals in mind, each one having its own set of requirements.

\begin{enumerate}
    \item Biological shielding, to enable personnel access as quickly as possible after beam shutoff.  While the beam is running radiation levels will be too high for personnel access, regardless of shielding design.  But it is important to minimize short-lived activity of items in the IsoDAR caverns, considered in the sections  \ref{components} and \ref{tritium} on component activation and tritium production.
    \item Rock activation, as neutrons that penetrate the rock surfaces can cause activation, and some long-lived isotopes, such as $^{22}{\rm Na}$ may become a problem if activity persists beyond the lifetime and decommissioning of the experiment (Section \ref{rock}.)
    \item Neutrons or gammas leaking into the fiducial volume of the liquid scintillator detector.  Here the goal is to keep these fluxes below the natural background levels one would see in the detector. (sections \ref{detector} and \ref{photons}.)
\end{enumerate}

In the studies for siting IsoDAR at the Kamioka Observatory, the high sodium content of the Inishi rock at this site dominated the shielding requirements~\cite{Shieldingpaper}, and led to the very large block we will describe.

Figure~\ref{fig:shielding} shows our designed shielding configuration. The roughly-spherical sleeve is enclosed in a solid iron cylinder (aligned with the beam direction) with an outer lateral diameter of 340~cm. The long axis of the cylinder is also 340 cm.  This cylinder is surrounded by boron-loaded concrete, poured into a form that is a cube 520 cm on a side. 
The minimum concrete thickness is 90 cm along the axes of the cube, and capping the ends of the cylinder.  
These thicknesses were found to be adequate in a previous study for the KamLAND site~\cite{Shieldingpaper}, 
and represent very conservative  limits for deployment at Yemilab.
The simulation geometry includes a further concrete plug on the right of the figure, in the direction away from the beam entry (which points away from the detector) to block fast neutrons transported down the torpedo cooling pipes from reaching the cavern, and a further iron block at the beam entry, on the side pointing towards the detector, which represents the final bending magnet on the beamline. The simulated outgoing neutron fluxes are recorded on a notional sphere of radius 460 cm around this system. 
Note, this sphere marks the coordinates of a neutron crossing its surface, but does not record the angle at which it crosses. 

In addressing the various  topics enumerated above related to shielding and radiation levels in the experimental areas, we present primarily the methodology for addressing these topics.  
For each topic we collect 
the relevant appropriate data where available (e.g. the composition of the rock)
or clearly state the assumptions we are making where it is not (e.g. the nature of the accelerator components). We use the Geant4 simulations to quantify possible problems, and whether the mitigations we have made are adequate. 
As stated earlier, our calculations relate specifically to the Yemilab site, however they can be tailored to apply to any experimental site by inserting appropriate local conditions.

\subsubsection{Local Backgrounds}

\begin{figure}[t!]
    \centering
    \includegraphics[width=1.0 \columnwidth]{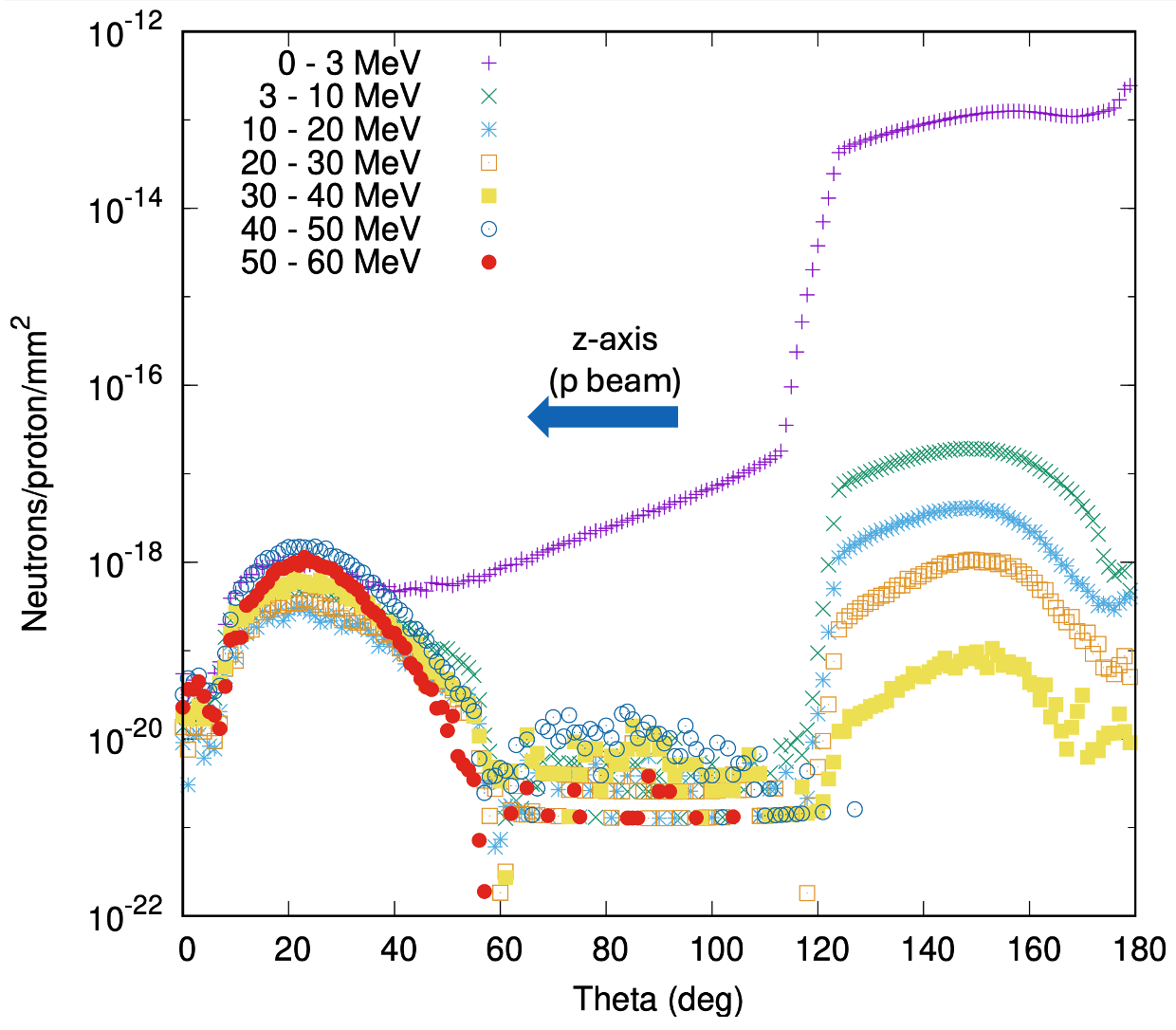}
    \caption{Neutron number density emerging from the shielding, using a ${\rm Be-D_{2}O}$ target. See text for details. }
    \label{fig:flux}
\end{figure}

Figure~\ref{fig:flux} shows the 
neutron density per incident proton resulting from the shielding arrangement in Figure~\ref{fig:shielding} for various neutron energies as a function of  $\theta$, the polar angle  with respect to the proton direction, recorded on a sphere of radius 460 cm. The  units are neutrons per square mm (on the notional sphere) per incident proton: to convert to the perhaps more familiar units of ${\rm cm}^{-2}{\rm s}^{-1}$ they can be multiplied by $6.24 \times 10^{18}$.\footnote {A typical flux in a large power reactor is of order $10^{13} n\ {\rm cm}^{-2}{\rm s}^{-1}$.}

At high energies the neutrons peak in the forward direction, which, because of the $180^\circ$ bend in the beamline, is away from the detector.
There is a comparatively large flux at large angles, heading towards the detector. These are low energy neutrons emerging from the back face, where the beam hits the target, and going backwards down the beam pipe, perhaps scattering a few times on the way.  Neutrons are not stopped by the iron in the final bending magnet, because iron is more or less transparent to neutrons below 3 MeV.
 However, as mentioned earlier, in this study the neutrons below 3 MeV are disregarded anyway, because they do not lead to background events in the detector. 

\begin{figure}[t!]
    \centering
    \includegraphics[width=0.5 \columnwidth]{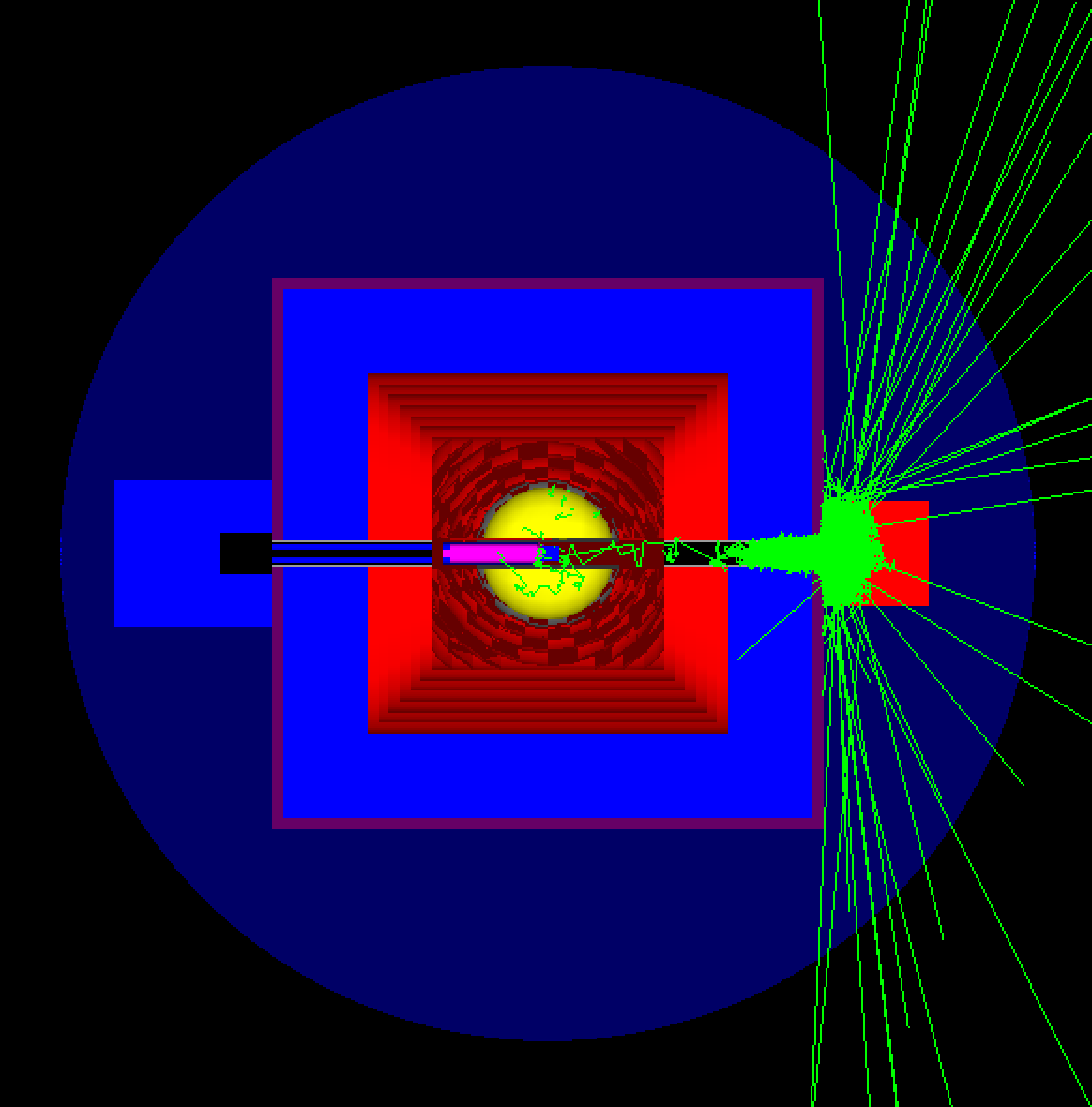}
    \caption{A few typical neutron trajectories from the target. }
    \label{fig:fluxexplain}
\end{figure}

 The step around 120$^\circ$  is produced by neutrons emerging from the pipe at $z \approx -260$ cm that can then reach the recording sphere.
 This feature of the plot 
 can be understood by considering a few typical 
 neutron trajectories, as shown in Figure~\ref{fig:fluxexplain}. Note the beam direction matches Figure~\ref{fig:flux} but is the opposite of Figure~\ref{fig:shielding}.

\subsubsection{Activation of components}{\label{components}}

Activation of the accelerator components due to beam loss along the transport beam line and neutrons emerging from the target and shielding is a common problem for accelerators. 
However, another source of activation comes from neutrons emerging from the shielding.  While beam loss can be addressed by careful optics, and controlled loss points along the beamline (collimators surrounded by good shielding), the background of neutrons from the shielding cannot be so controlled.
The principal activation problem will come from copper in the cables and magnet windings. 
To estimate the magnitude of this issue, simulations were 
done by placing a 1 cm thick Cu plate on the upper surface of the shielding block.  
Only gamma emitting isotopes with a lifetime longer than 24 hours were recorded in the simulation.

A radionuclide inventory shows that the isotopes that are the main contributors are  $^{64}{\rm Cu}$,  $^{57}{\rm Co}$,  $^{58}{\rm Co}$,  $^{60}{\rm Co}$ (Table \ref{tab:table1}) although other isotopes like 
$^{56}{\rm Co}$, $^{51}{\rm Cr}$, etc are also present but at much lower rates.

\begin{table}[b]
\caption{List of the main radio-isotopes inside copper cables.}
\begin{center}
\begin{tabular}{|c |c |c|c|} 
 \hline\hline
 \bf{Isotope} & \bf{Half life} & \bf{Isotope/proton} & \bf{Decay mode}\\[0.5ex] 
 \hline
 $ ^{64} {\rm Cu}$ & 12 h & 1.88e-16& $\beta^{+}$, $\beta^{-}$\\ 
 \hline
  $^{57} {\rm Co}$ & 270 d & 2.25e-17 & $\epsilon$\\
 \hline
  $^{58} {\rm Co}$ &  70 d & 6.01e-17 & $\beta^{+}$\\
 \hline
 $^{60}{\rm Co}$ & 1900 d & 3.76e-17 & $\beta^{-}$\\
 \hline
 \hline
\end{tabular}
\end{center}
\label{tab:table1}
\end{table}

\begin{figure}[b!]
    \centering
    \includegraphics[width=1.0 \columnwidth]{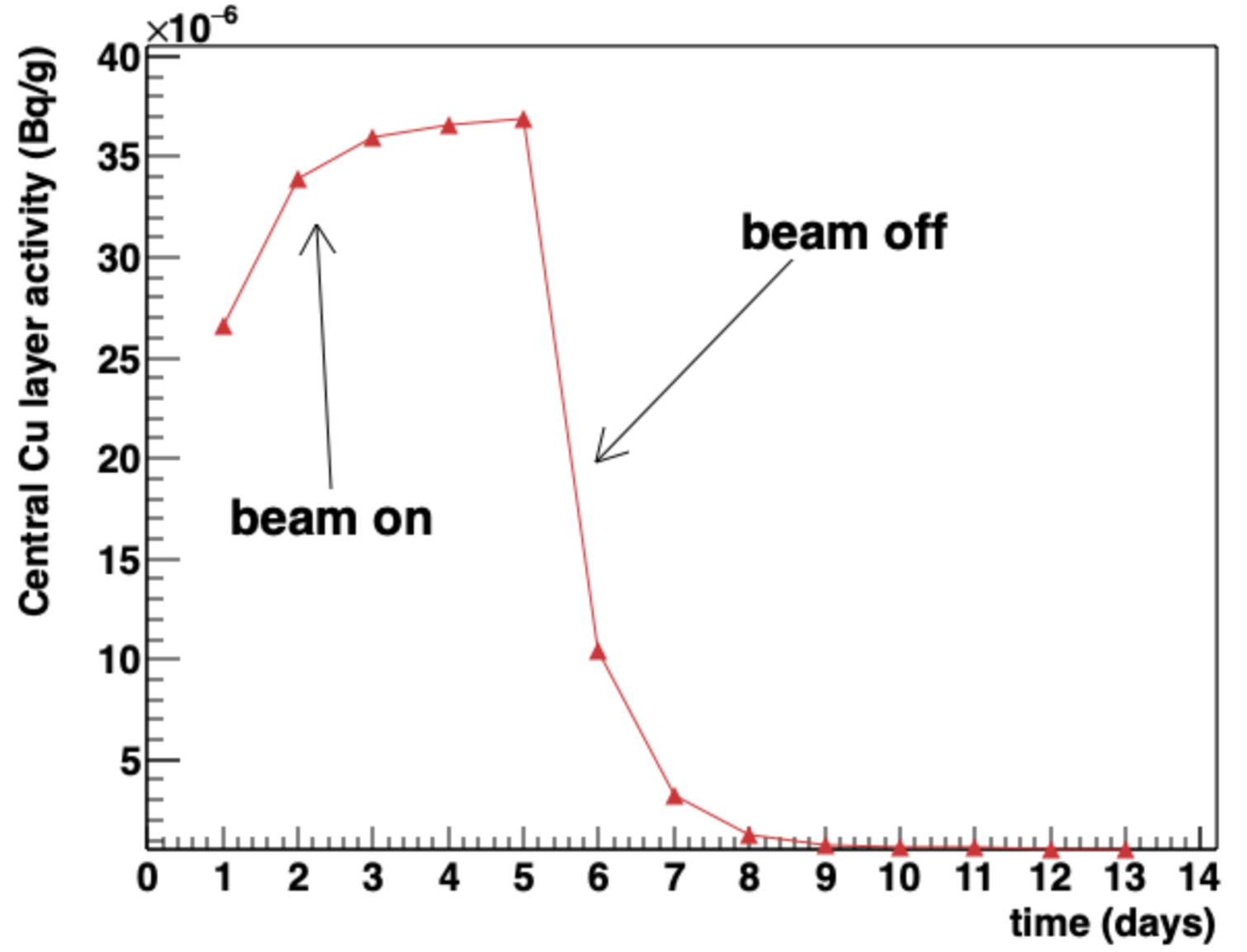}
    \caption{Induced activity in a copper plate placed above the IsoDAR target, sleeve, and shielding block.}
    \label{fig:copper}
\end{figure}

The activity of the Cu plate from this simulation is shown in Fig.~\ref{fig:copper}. During beam on operation, additional radiation comes from beam loss (neutrons, $\rm {gamma}$ rays) or from electronic equipment (X-rays from RF amplifiers). During operation no personnel will be allowed in the area. When the beam is off, there will be a small amount of radiation from activated material which will be minimal  after several days. 

Our 60 MeV protons are substantially lower in energy than beams at CERN or Fermilab.  In these areas, spallation reactions with very high energy neutrons will activate the air with short lived neutron-deficient isotopes of nitrogen or other light elements, including $^7{\rm Be}$.  These can present significant environmental hazards to personnel, but as stated our energies are too low to produce these activities.  
Nonetheless, in the spirit of overcautiousness, IsoDAR will plan on keeping its environment at a slight negative pressure, so air from the main (clean) laboratory will always flow into the IsoDAR area, from where it will be pumped directly to the surface.

\subsubsection{Activation in the rock}{\label{rock}}

Some neutrons will escape the shielding during the experiment and will be absorbed in the rock causing rock activation. 
On completion of the experiment
the resulting residual radiation must be low enough to enable the cavern to be open to other uses.  

This depends not only on the escaping neutron flux but on the composition of the surrounding rock and also on the legal definition of ``low enough".
We have compared a possible installation at Yemilab in Korea~\cite{Yemilab} with our previous study 
which assumed installation at KamLAND in Japan~\cite{Kamland}.

The regulation for the allowed rate at Yemilab is 10 Bq/g, a factor of 100 bigger than at KamLAND (0.1 Bq/g).
An assay of Yemilab rock showed that various types of rocks were found, limestone being the most abundant (almost 90\% pure ${\rm CaCO}_{3}$) followed by quartz porphyry, pyrite and skarn minerals (silicate rocks which contain Ca, Fe, Mn, Al and Si minerals). Unlike the Inishi type rock found at KamLAND, the Yemilab environment is a factor of 10 lower in critical elements, with low concentrations of Eu ($< 1$~ppm ), Co ($< 10$~ppm) and Na ($0.02\%$). The measured rock composition can be found in Table~\ref{tab:table2}.

\begin{table}[t]

\caption{The rock composition at Yemilab.}
\begin{center}
\begin{tabular}{|c c c c |} 
 \hline\hline
 \bf{Compound} & Composition & \bf{Compound} & Composition \\ [0.5ex] 
 \hline
 $\bf{CaCO_{3}}$ & 90\% & \bf{U} & 0.8 ppm \\ 
 \hline
 \bf{Na} & 0.022\% & \bf{Th} & 3.3 ppm\\
 \hline
 \bf{Co} & 6 ppm & \bf{K} & 11800 ppm \\
 \hline
 \bf{Eu} & 0.2 ppm &  &  \\
 \hline
 \hline
\end{tabular}
\end{center}
\label{tab:table2}
\end{table}

 The activation level will be  measured when the experiment will be dismantled (about 4-5 years after the beam is turned off) and any rock activation of isotopes with half lives of less than one year or so would have totally decayed away. The rock activation was estimated using the same technique as used for KamLAND~\cite{Shieldingpaper} but with the Yemilab rock composition. The long lived isotopes ($\tau$ $>$ 2-3 years) like $^{22} {\rm Na}$, $^{60} {\rm Co}$ and the $\rm Eu$ pair present in the rock are activated by neutrons penetrating the rock. $^{22}$Na is produced by fast neutrons via the (n,2n) channel (11 MeV threshold), while the others are produced by thermal neutron activation.
Table~\ref{tab:table3} shows a comparison between the abundance of the progenitor isotopes at KamLAND and Yemilab.

\begin{table}[b]
\caption{Long lived isotopes comparison at Yemilab and KamLAND. }
\begin{center}
\begin{tabular}{|c c c c c |} 
 \hline\hline
 \bf{Isotope} & \bf{Half life} & \bf{Parent} & \bf{KamLAND} & \bf {Yemilab} \\ [0.5ex] 
 \hline
 $^{22} {\rm Na}$ & 2.6 y & $^{23} {\rm Na}$& 6\% & 0.022\% \\ 
 \hline
  $^{60} {\rm Co}$ & 5.3 y & $^{59} {\rm Co}$ & 30 ppm & 6 ppm\\
 \hline
 $^{152} {\rm Eu}$ & 13.5 y & $^{151} {\rm Eu}$ & 3 ppm & 0.1 ppm\\
 \hline
 $^{154} {\rm Eu}$ & 8.6 y  & $^{153} {\rm Eu}$  & 3 ppm & 0.1 ppm\\
 \hline
 \hline
\end{tabular}
\end{center}

\label{tab:table3}
\end{table}

\begin{figure}[t!]
    \centering
    \includegraphics[width=1.0 \columnwidth]{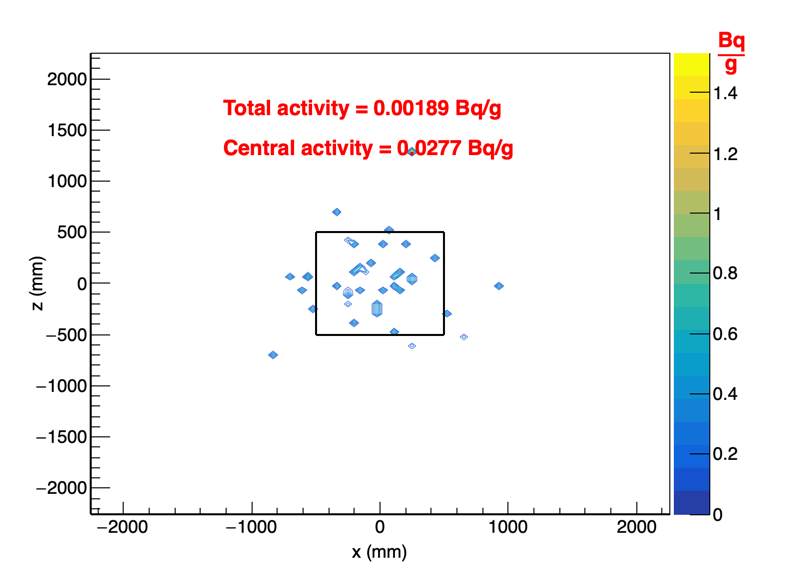}
    \caption{Spatial distribution of induced activity in the first 10 cm of the rock layer, closest to the shielding block (Fig.~\ref{fig:shielding}). The induced activity was calculated in Bq/g. In this image the x-z plane is parallel with the top surface of the shielding block.}
    \label{fig:hot-spot}
\end{figure}

The simulations included a block of rock placed above the shielding with a 100~cm thickness to evaluate the induced activity at various depths in the rock. The total activity was calculated over the entire rock volume, however the highest activity concentration was measured on the central hot spot. Preliminary studies have shown that the induced activity averaged over a central 1 $m^{2}$ region is 0.028 Bq/g. Even for a shielding configuration consisting of 30 cm Fe and 90 cm of boron loaded concrete the highest central activity was less than $0.05$~Bq/g (Figure~\ref{fig:hot-spot}). This is extremely low compared to the Korean legal requirement ($10$~Bq/g) and proves that rock activation at Yemilab is not an issue. This favorable result is due to several factors. First, the sodium concentration in the rock is very low, as is the concentration of 
europium. In addition, the cobalt contamination in the rock is 5 times lower at Yemilab.

\subsubsection{Tritium production}{\label{tritium}}

Tritium is produced in the target system in both the sleeve (by interaction of neutrons with Li) and heavy water coolant. 
It is a radiation hazard because of its long half life ($\approx$12 years) and the emission of beta particles.
The tritium produced in the sleeve will remain trapped inside and will not be an issue, however the tritium produced in the coolant could be a problem because of potential leaks.
It was found that the tritium produced in the heavy water with the Yemilab design is almost four times less than for the KamLAND version, and the exact values per incident proton in both materials are shown in Table \ref{tab:tritium}.
These differences result from changes in the target and sleeve geometries.

Tritium production could be a major issue in environments with difficult-to-control ground water in the experimental area.  This was a serious issue at KamLAND, but fortunately is nonexistent at Yemilab, which is dry.  Generally, mitigation of tritium generated in ground-water must be conscientiously addressed because of the difficulty in containment of this contaminant, and its potential for entering the local aquifers.

\begin{table}[t]
\caption{The tritium production in the Yemilab and KamLAND target/sleeve systems.}
\begin{center}
\begin{tabular}{|c c c  |} 
 \hline\hline
 \bf{Material} & Yemilab (T/POT) &  KamLAND (T/POT)\\ [0.5ex] 
 \hline
 coolant & $3\times10^{-5}$& $11\times10^{-5}$ \\
 \hline
 sleeve & $36.4\times10^{-3}$& $1.7\times10^{-3}$ \\
 \hline
 \hline
\end{tabular}
\end{center}
\label{tab:tritium}
\end{table}

\subsubsection{Neutron Background in the detector}{\label{detector}}

Neutrons produced by proton interactions in the target system can be a background to analyses if they reach the detector. The physics analyses will require a $>3$~MeV signal to reduce confusion with neutron capture, but a fast neutron signal may be above this threshold. Neutron background can be reduced by pulse shape identification that discriminates between neutron-induced protons and electrons/positrons.  Nevertheless, it is best to minimise the intrinsic neutron rate into the detector rather than rely on cuts.

\begin{figure}[t!]
    \centering
    \includegraphics[width=1.0 \columnwidth]{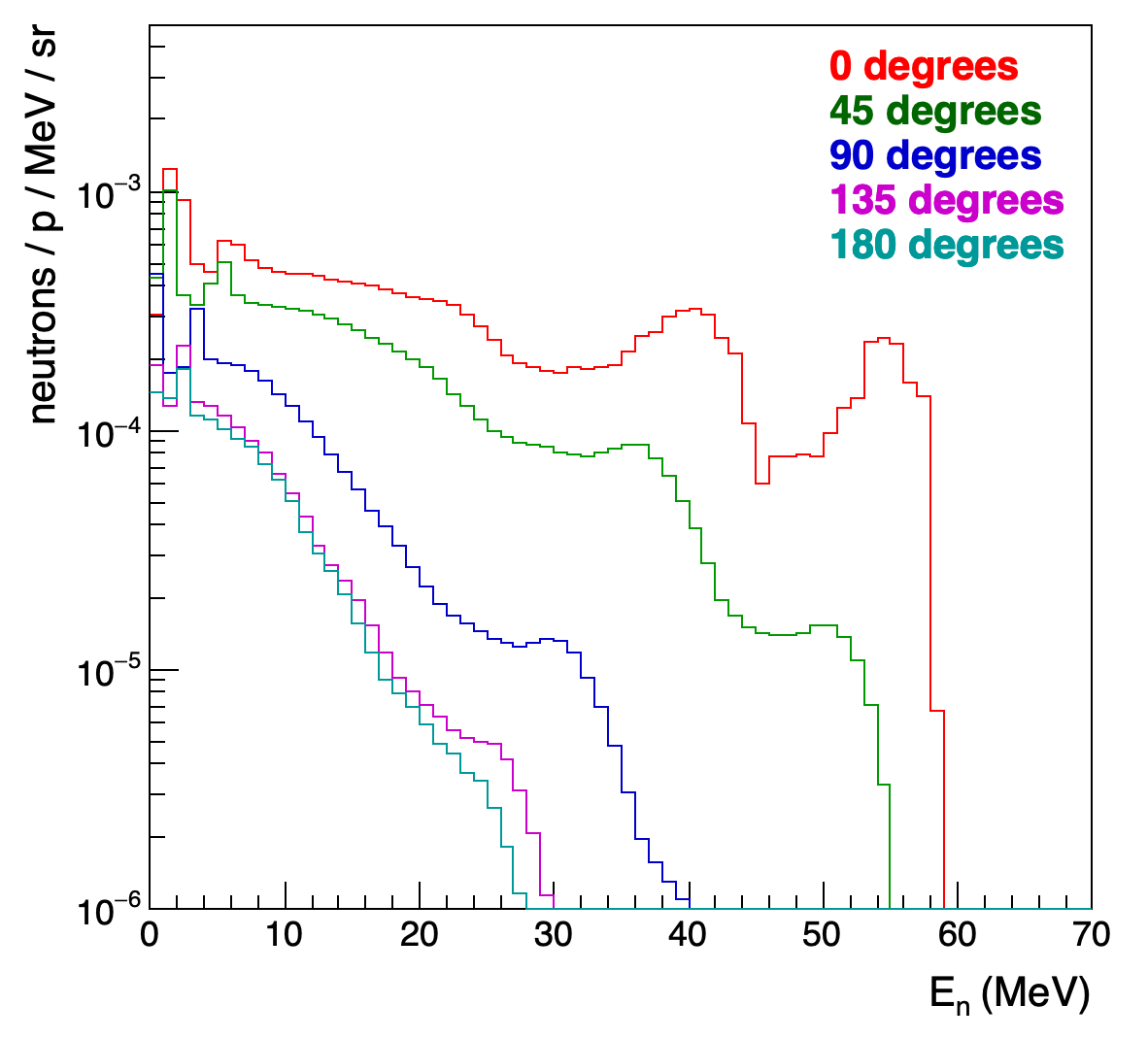}
    \caption{Neutron spectra out of target for various angles.}
    \label{fig:n-out-target}
\end{figure}

At KamLAND, the shielding consisted of a combination of 2 m Fe and B-rich concrete placed around the target system with an extra 2 m length  additional shielding of Fe towards the detector to further shield and prevent the neutrons from entering the detector~\cite{Shieldingpaper}. While at KamLAND, the shielding design was constrained by the limited space of the cavern, this is not the case at Yemilab where the room that will accommodate the target and shielding system is quite large (7 m wide, 7 m maximum height at the peak of a semicircular dome with radius 3.5 meters, and 22 m long). The energy spectra of the neutrons produced by the target for different angles with respect to the incident proton beam are shown in Figure~\ref{fig:n-out-target}. The backwards-going neutron flux is greatly reduced and this accounts for the target orientation as fast neutrons must not be allowed to reach the detector (note that the beam undergoes two $90^{\circ}$ bends so that it hits the target while moving away from the detector).

\begin{figure}[b!]
    \centering
    \includegraphics[width=1.0 \columnwidth]{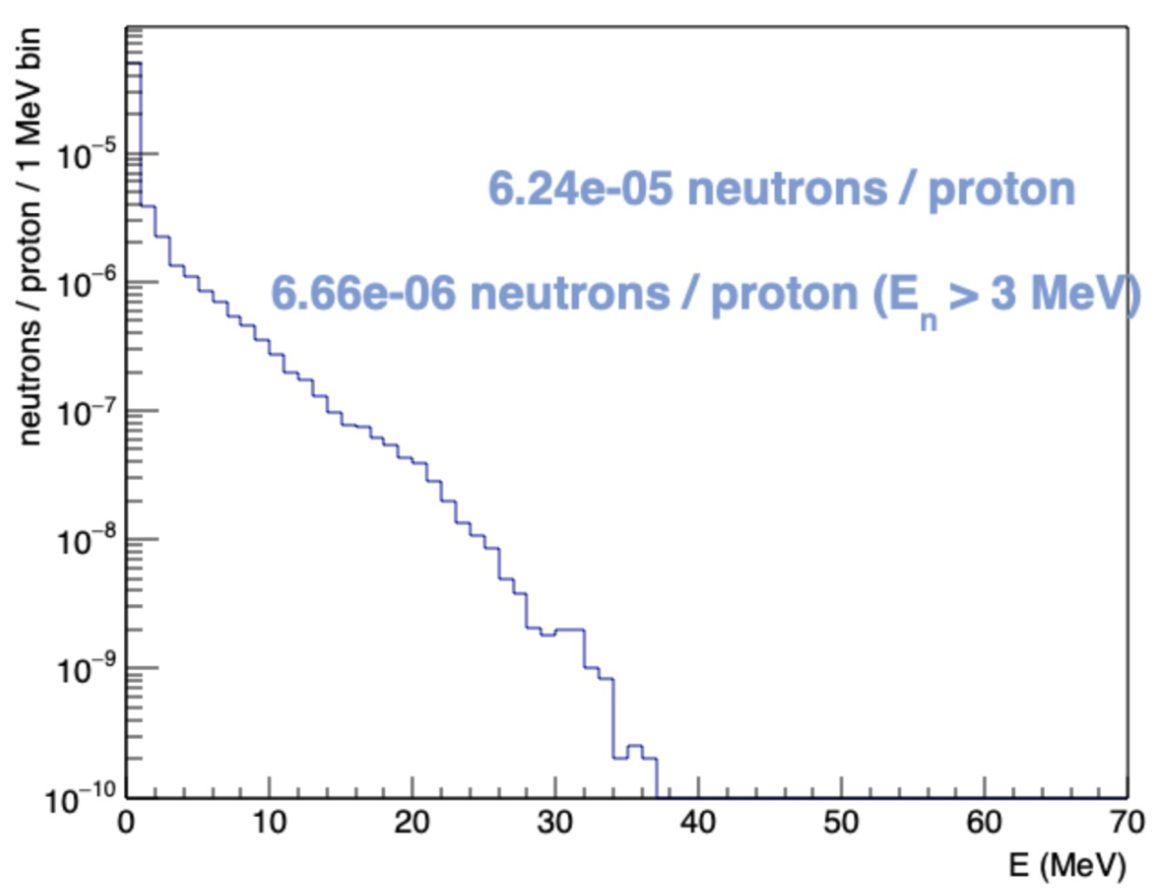}
    \caption{Neutron spectrum out of the shielding through the proton beam pipe.} 
    \label{fig:n-out-beam-pipe}
\end{figure}

The energy spectra of the neutrons going through the vacuum pipe and 
emanating at the surface of the shielding in the backward direction is shown in Figure~\ref{fig:n-out-beam-pipe}.  To estimate the neutron background reaching the detector, these neutrons were tracked through an additional 5 metres thick iron block (a nominal 1 m for the last bending magnet and then an additional 4 m) , since it is the fast (above 3 MeV) neutrons that can contribute to the additional background. Iron was the material of choice due to its high effective cross section for removing high energy neutrons. These neutrons spread out inside the block of iron, and their intensity is effectively attenuated. 

The picking of 5 meters is not that we view this as our thickness requirement, but as a gauge of whether or not the required attenuation can be achieved at all.  As we see, the answer is a resounding Yes, allowing then the optimization of the shielding thickness to match the estimated (or measured) natural flux of fast neutrons at the Yemilab site.

Ideally, the shielding to attenuate these neutrons should reduce the rate in the fiducial volume to below the naturally-occurring neutron level, which results from nuclear reactions induced by high-energy muons that reach the environment of the detector and from natural radioactivity in the environment.  To the best of our understanding, this is ``a few neutrons per year (above 3 MeV) in the fiducial volume.''  This represents, consulting the numbers shown in Fig.~\ref{fig:n-out-beam-pipe}, an overall attenuation factor of about 10$^{-24}$.

\begin{figure}[t!]
    \centering
    \includegraphics[width=\columnwidth]{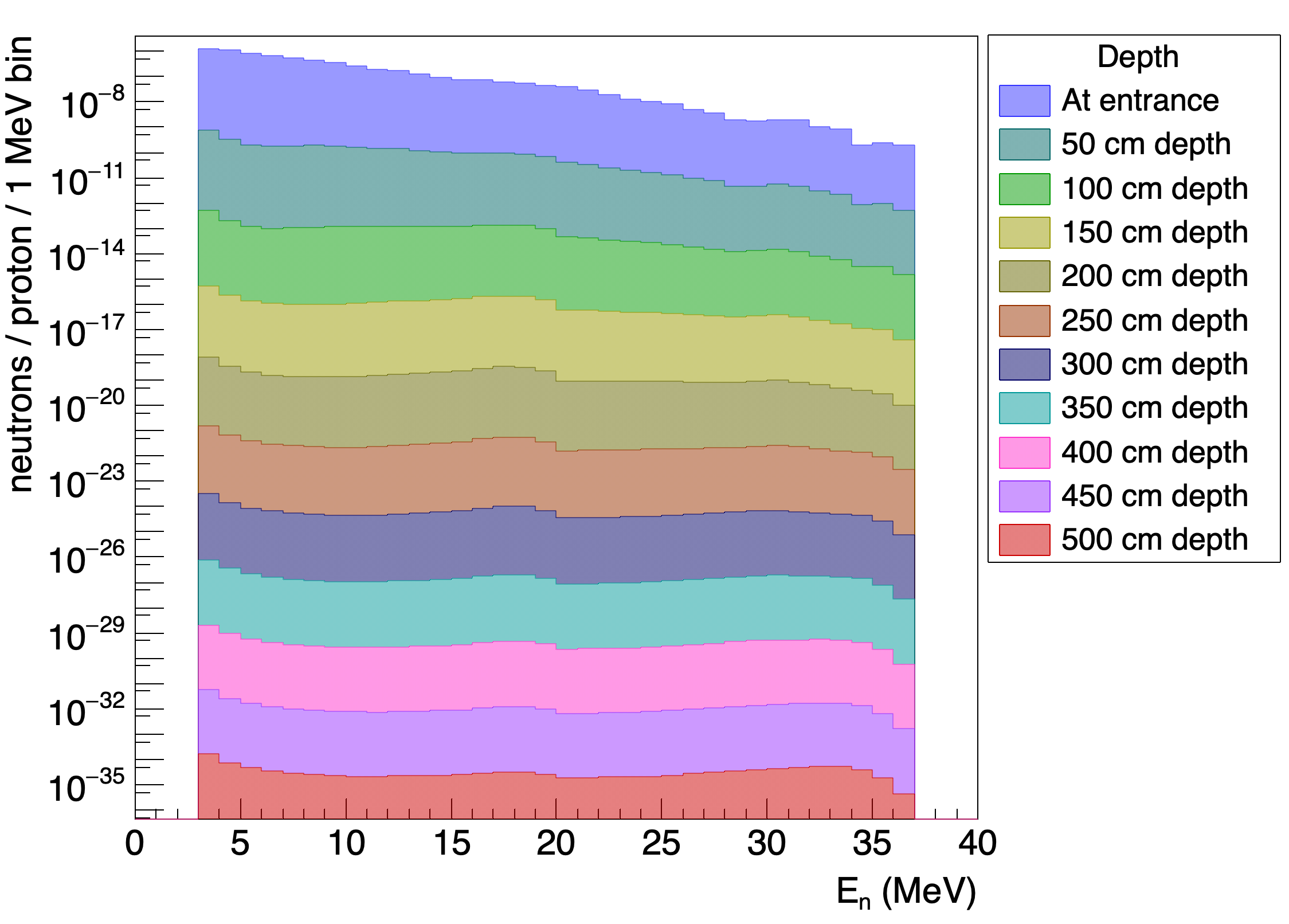}
    \caption{Neutron spectra at various depths inside the 5 m of iron shielding.}
    \label{fig:n-spectra}
\end{figure}

The same technique used for KamLAND simulations, event biasing using particle splitting, was applied to evaluate the neutron flux towards the detector. However, the 5 metres of iron was far too great even when using these variance reduction techniques, and therefore this simulation was done in consecutive stages. The 5 metres block was divided into 10 layers, 50 cm thick. The energy spectrum and spatial distribution of neutrons were recorded at the end of each layer, and a new simulation was started using these values as input parameters for the following Fe layer. For a correct normalisation of the results, after each stage the results were scaled by a factor equal to the ratio between the number of neutrons per proton that were expected to enter the new shielding layer and the number of neutrons generated in the simulation to produce results with good enough statistics. This has enabled the simulation of the entire thick block of iron, and
we find that 500 cm reduces the flux of neutrons  by a factor $1.60 \times 10^{-28}$. (The use of consecutive stages in the simulation reduces the statistical error on this to $1.84 \times 10^{-31}$.) 
The energy spectra of the neutrons recorded after each iron layer are shown in Figure~\ref{fig:n-spectra}. 
(The origin of the step around 20 MeV in many spectra is unclear: it may or may not be related to the switch between measurements and models in the Shielding Physics List.)
The 3 MeV energy cut was applied in all these simulations.

\begin{figure}[t!]
    \centering
    \includegraphics[width=\columnwidth]{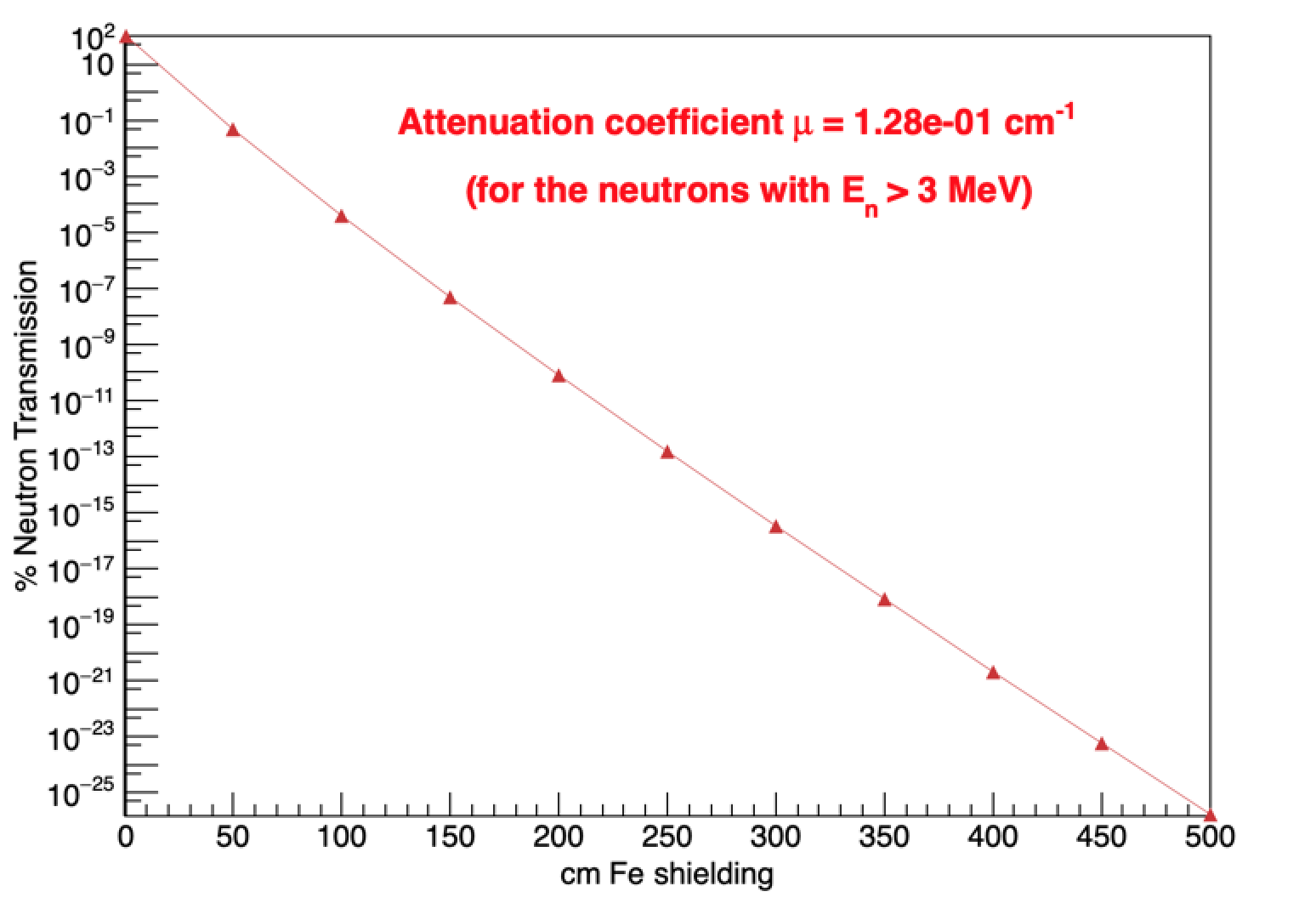}
    \caption{Neutron transmission through the shielding.}
    \label{fig:n-transmission}
\end{figure}

5 m of Fe reduces the neutron rate to $\approx 10^{-35}$ neutrons/proton
. This value is much lower than was obtained for the KamLAND shielding. The neutron attenuation in iron recorded in these simulations is shown in Figure~\ref{fig:n-transmission}. For the single-flash elastic scattering channel we anticipate a signal rate of $\sim 7000$ events over 4 years~\cite{Alonso:2021kyu}(the double-flash IBD signal size is much larger) or $\sim 7\times 10^{-23}$ events per proton, so this shielding is more than adequate to render neutron-induced background negligible.

\begin{figure}[t!]
    \centering
    \includegraphics[width=14 cm]{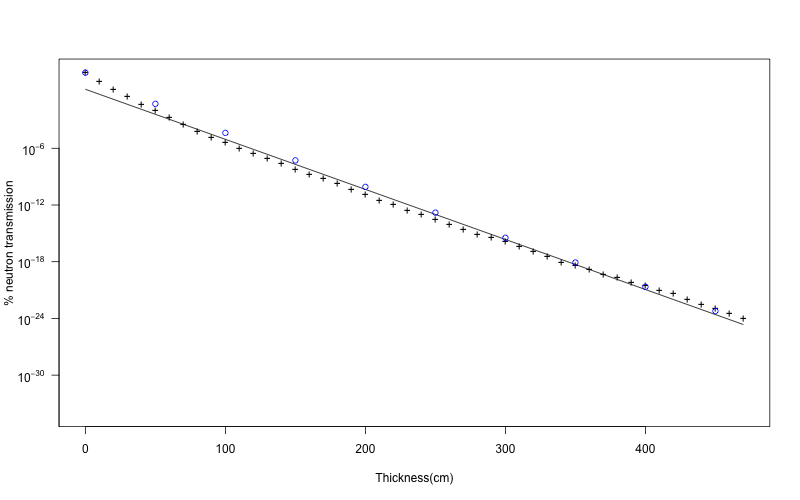}
    
    \caption{Neutron transmission through the shielding according to MCNPX (black crosses), showing the best exponential fit $\propto e^{-0.18 x}$ (black line). The blue circles are the Geant4 results shown in Figure~\ref{fig:n-transmission}, for comparison, both being normalized to 100\% at zero thickness. }
    \label{fig:MCNPa}
 \end{figure}
 
    

As the attenuation is so large and the result so important, we have done a simple cross check of our Geant4 results against those of the program MCNPX~\cite{MCNP}. For this purpose
the full geometry was not modelled: 60 MeV protons were generated in a beryllium slab, creating neutrons.
The shielding was represented by successive square iron sheets, 300 cm a side and 10 cm thick, encased in a large cylinder of calcium carbonate.
The sampling importance of successive 10 cm sheets increased by a factor of 3.0 
which compensates for the attenuation and thus maintains the statistical precision (neutrons crossing the boundary are replaced by 3 neutrons with 1/3 of the weight). Results are shown in Fig.~\ref{fig:MCNPa}. The attenuation is roughly (though not exactly) exponential with an attenuation coefficient $0.1811 \pm 0.0007 {\rm cm}^{-1}$,where the error is estimated 
from subsampling.
This attenuation coefficient is similar but not identical to that from Geant4; in this comparison it should be noted that the MCNPX geometrical model is simpler and that the attenuation is not an exact exponential so a difference is to be expected. Also the MCNPX result depends on the weight multiplier used: a factor of 2.0 is understandably not sufficient, factors of 4.0 or 5.0 predict greater attenuation at large thicknesses. This feature of the MCNP code is not understood, but it probably due to particle weights falling below $1.18 \times 10^{-38}$, the 
smallest normal single-precision floating point number. In summary, the MCNPX simulation agrees broadly with our Geant4 model, and where it disagrees it does so by predicting even greater attenuation.

\subsubsection{Photon background in the detector}{\label{photons}}

There will also be a background due to gamma rays, produced  mainly by inelastic processes and also, at much lower rates, by neutron capture and subsequent gamma decay of the excited isotope formed.
Photon inelastic processes and radioactive decay contribute even less. Other sources of photon background include beam loss along the transport line due to gas interactions or loss at collimators or the walls of the beam pipe. 

This photon background is not relevant 
for the detection of $\overline{\nu}_e$
particles through IBD events, as these use the ``double flash'' coincidence of the
prompt positron and neutron capture events, but could
in principle be a background for BSM
searches which use a ``single flash" signal
from (e.g.) $\overline{\nu}_e-e^-$ elastic scattering.

Studies show \cite{Alonso:2021kyu} that 
the elastic scattering process has backgrounds from
solar neutrinos and from IBD events with
one of the `flashes' undetected, with smaller contributions induced by cosmic rays and radiogenics.  Mitigating these requires a
visible energy threshold of $>$3 MeV,
and a signal size of $\sim$7000 $\overline{\nu}_e-e^-$ events (over 5 years running) is anticipated.

\section{Discussion and conclusions}

The design of the IsoDAR target and the sleeve 
has evolved, with increased understanding of the challenges 
of target design required to cool the 600~kW target
in a confined space.
Engineering details are presented in a separate publication\cite{PDR}.  
The simulations 
reported in this paper show that these challenges can be met without compromising the physics reach of the experiment,
although high isotopic
purity will be required for the lithium in the sleeve.

We report on the development of simulation tools to address a number of issues related to operating an experiment generating high levels of radiation in a low-background laboratory, specifically:
\begin{enumerate}
\item Optimisation of primary shielding design around the target
\item Activation of copper components
\item Tritium production
\item Activation of the rock 
\item Neutron backgrounds in the detector
\item Photon backgrounds in the detector
\end{enumerate}
The results obtained 
encourage us to believe
that such experiments can be successfully deployed without affecting the sensitivity of other experiments in the underground laboratory.

\section{Acknowledgements}

We are grateful to our colleagues from the Center for Underground Physics and Yemilab, 
in the Republic of Korea who have provided us with detailed descriptions of the 
relevant conditions in their laboratory.

This work was supported by the US National Science Foundation through grants PHY-1912764 and PHY-1707969. DW was supported by the Heising-Simons foundation.

\bibliographystyle{JHEP}  
\bibliography{refs.bib}
\end{document}